\documentclass[twocolumn,prl,nofootinbib,preprintnumbers,tightenlines,superscriptaddress]{revtex4}

\pdfoutput=1
\usepackage[english]{babel}
\usepackage{amsmath,amssymb,amsfonts}
\usepackage{graphicx}
\usepackage[normalem]{ulem}
\usepackage{slashed}
\usepackage{booktabs}
\usepackage{color}
\usepackage{braket}
\usepackage[sort&compress]{natbib}
\usepackage{tikz}

\usepackage[
	colorlinks=true,
	citecolor=black,
	linkcolor=black,
	urlcolor=blue,
	hypertexnames=false]{hyperref}
\usepackage{cleveref}

\newcounter{qnumber}

\newcommand{\arXiv}[2]{\href{http://arxiv.org/pdf/#1}{{\tt #2/#1}}}
\newcommand{\arXivold}[1]{\href{http://arxiv.org/pdf/#1}{{\tt #1}}}

\newcommand{\beq}{\begin{equation}}
\newcommand{\eeq}{\end{equation}}

\newcommand{\colvec}[1]{\left(\begin{array}{c}#1\end{array}\right)}

\newcommand{\bk}[1]{\left<#1\right>}
\newcommand{\bks}[1]{\left[#1\right]}

\newcommand{\sket}[1]{\left|#1\right]}
\newcommand{\sbra}[1]{\left[#1\right|}

\begin{document}

\title{Monopoles Entangle Fermions}

\author{Csaba Cs\'aki}
\email{ccsaki@gmail.com}
\affiliation{Department of Physics, LEPP, Cornell University, Ithaca, NY 14853, USA}

\author{Yuri Shirman}
\email{yshirman@uci.edu }
\affiliation{Department of Physics $\&$ Astronomy, University of California, Irvine, CA 92697, USA}

\author{Ofri Telem}
\email{t10ofrit@gmail.com}
\affiliation{Department of Physics, University of California, Berkeley, CA 94720, USA}
\affiliation{Ernest Orlando Lawrence Berkeley National Laboratory, Berkeley, CA 94720, USA}

\author{John Terning}
\email{jterning@gmail.com}
\affiliation{QMAP, Department of Physics, University of California, Davis, CA 95616, USA}

\begin{abstract} 
We resolve the decades old mystery of what happens when a positron scatters off a minimal GUT monopole in an s-wave, first discussed by Callan in 1983. Using the language of on-shell amplitudes and pairwise helicity we show that the final state contains two up quarks and a down quark entangled with angular momentum stored in the gauge fields, which is the only particle final state that satisfies angular momentum and gauge charge conservation. The cross section for this process is as large as in the original Rubakov-Callan effect, only suppressed by the QCD scale. The final state we find cannot be seen in Callan's truncated 2D theory, since our entanglement requires more than 2 dimensions.
\end{abstract}

\maketitle

\section{introduction}
The scattering of electrically charged fermions with magnetic monopoles is a very peculiar process \cite{reviews}. Until recently the theoretical understanding of these processes faced three major difficulties: 
\begin{itemize}
\item Weinberg \cite{Weinberg:1965rz} found that the scattering amplitudes are not  Lorentz invariant.
\item Multiparticle scattering states with both electric and magnetic charges carry additional angular momentum in the gauge field \cite{Thomson,ZwanzigerSelection} and cannot be written as tensor products of Wigner's one-particle states.
\item The analysis of the scattering of Grand Unified Theory (GUT) monopoles seemingly led to the conclusion that one must either give up on conservation of gauge charges or accept the existence of fractional particles \cite{Callan:smatrix}. 
\end{itemize}
The Lorentz violation problem was resolved by all orders resummation in  \cite{Terning:2018udc} and order-by-order for the special case of monopoles bound with anti-monopoles in \cite{Terning:2020dzg}. The problem with multiparticle states was resolved by the inclusion of an additional quantum number called pairwise helicity \cite{Csaki:2020inw,Csaki:2020yei}. In this letter we will present a resolution of the final problem.

To understand the essence of the final problem regarding the scattering of GUT monopoles it is helpful to recall that in a $U(1)$ theory with a magnetic monopole and two oppositely charged massless Weyl fermions,  angular momentum conservation implies that there can be no forward scattering in the $J=0$ partial wave: the massless fermion must flip its chirality~\cite{Kazama} by turning into the CP conjugate of the other fermion. We can embed this simple theory into a `t Hooft-Polyakov model \cite{'tHooft:1974qc} with an $SU(2)$ gauge group and two Weyl doublets\footnote{An even number of doublets is required to avoid the Witten global anomaly.}. This theory has an $SU(2)$ flavor symmetry, which is perfectly consistent with the helicity flip process if the flavor flips as well. Things become more subtle in a model with 4 doublets where the helicity flip process is forbidden by an $SU(4)$ global symmetry.  The global symmetry allows processes with one fermion initial state scattering into three fermion final states as well as processes where \cite{Callan:smatrix}
two incoming fermions to scatter to two outgoing fermions \cite{RubakovCallan}. The proton decay catalized by the latter type of process produce the leading observational bounds on the relic density of GUT monopoles in the universe \cite{bounds}.

Callan \cite{Callan:smatrix} was the first to study the scattering of a positron  off a GUT monopole. The problem reduces to the four flavor `t Hooft-Polyakov model in a limit where some gauge couplings are dropped, implying again that there have to be at least 3 fermions in the final state.  Truncating to 2D and reformulating the problem in terms of solitons, Callan concluded \cite{Callan:smatrix,Callan:1982au} that
there was no possible 3 fermion final state that preserved all of the gauge quantum numbers. As noticed by Witten, the truncated theory produced half-solitons (aka ``semitons") in the final state \cite{Callan:smatrix}, which they identified with  ``half particles" in the full theory. With the fermion masses set to zero these states, if they existed, would have to be true asymptotic final states far from the monopole, where perturbation theory can be reliably applied.  Since these states cannot arise in perturbation theory, Callan suggested \cite{Callan:smatrix} that there could be some kind of statistical understanding where charge conservation is violated in individual events but is conserved on average.  However this explanation was never fully embraced, since gauge invariance of the full 4D theory does not allow for such a probabilistic conservation.

The aim of this paper is to find the form of the amplitude for this process by making use of helicity amplitudes and pairwise helicity. We will indeed be able to identify the unique form of the amplitude that preserves all gauge quantum numbers, angular momentum and the approximate global flavor symmetry. As in the helicity-flip process discussed above, forward scattering is forbidden, and surprisingly the unique out state {\it entangles} the  three fermions  with the angular momentum in the gauge field, combining into a total $J=0$ partial wave. This amplitude corresponds to the lowest dimensional operator without derivatives in the low-energy effective theory. Thus we find that monopoles can produce entangled fermions with a cross section satisfying the s-wave unitarity bound. 

\section{Rubakov-Callan Interactions}

Consider the minimal $SU(5)$ GUT monopole obtained by embedding the standard 
`t Hooft-Polyakov monopole \cite{'tHooft:1974qc} into the $SU(2)_M\subset SU(5)$ generated by $T^i_{M}=\text{diag}\left(0,0,\tau^{i},0\right)$. Far away from the monopole, the full $SU(5)$ gauge symmetry is broken by a Higgs in the adjoint of $SU(5)$ to the standard model (SM) gauge group which includes the $U(1)_M$ generated by $T^3_M$. The monopole configuration is invariant under the combined rotations generated by $\vec{L}+\vec{T}$, where $\vec{L}$ is the orbital angular momentum operator and $\vec{T}$ is the vector of $SU(2)_M$ generators $T^i_{M}$.
In the $SU(5)$ GUT, every generation of SM fermions is embedded in a $\bar{\mathbf{5}}$ and a $\mathbf{10}$. Under $SU(2)_M$, these decompose into 4 doublets:
 \begin{eqnarray}\label{eq:doublets}
 \colvec{e\\-\bar{d}^3}\,,\,
 \colvec{\bar{u}^1\\u^{2}}\,,\,
\colvec{-\bar{u}^2\\u^{1}}\,,\,
\colvec{d^{3}\\\bar{e}}\,,
\end{eqnarray}
where the upper and lower components have charge $e_M=\pm\frac{1}{2}$ under $U(1)_M$, and all other fermions are $SU(2)_M$ singlets.
We labeled the particles by the corresponding left-handed fields, the right-handed particles correspond to hermitian conjugates of these fields. Here the $1,2,3$ label global color charge, which is broken in the vicinity of the monopole. Furthermore, in this paper we take the monopole to have charge $g_M=-1$ for consistency with Rubakov's notation \cite{Rubakov:1988aq}. This means that $q=e_M g_M=-1/2$ for $e,\bar{u}^1,\bar{u}^2,d^3$ while $q=1/2$ for $\bar{d}^3,u^2,u^1,\bar{e}$.
In the early '80s, Rubakov and Callan \cite{RubakovCallan} independently derived a remarkable feature of fermion-monopole scattering: when a pair of $u^{1}+u^{2}$ quarks is incident on the monopole, there are seemingly two possible outgoing states which conserve all SM quantum numbers - the initial state itself, and the state $d^{3^\dagger}+e^\dagger$. Rubakov and Callan showed that the $J=0$ part of the out state cannot, in fact, be the same as the in state, and so the entire $J=0$ partial wave is converted to $d^{3^\dagger}+e^\dagger$, thus violating baryon (B) number.

To see this effect, both Rubakov and Callan focused on a truncated theory in which one only retains the $J=0$ partial wave for each fermion. Famously \cite{RubakovCallan}, in this truncated theory, fermions with $q=-1/2$ exist only as incoming waves, while those with $q=1/2$ exist only as outgoing waves. In particular, this implies that forward scattering is always forbidden. Note that the hermitian conjugates of the fields in (\ref{eq:doublets}) have the opposite charge and helicity so $e^\dagger$ is also an outgoing wave.

Consequently, monopoles induce a B-violating process with a cross section saturating the $J=0$ unitarity bound. When QCD confinement is taken into account, this leads to monopoles catalyzing proton decay with a QCD scale cross section, counter to the naive intuition that the cross section is suppressed by the scattering energy over the GUT scale. The leading observational bounds on the relic density of GUT monopoles in the universe are then derived from proton/neutron decay catalysed by monopole capture in neutron stars \cite{bounds}.

\section{Monopole Catalysis and Pairwise Helicity}
In \cite{Csaki:2020inw}, the most general form of the S-matrix for the scattering of monopoles and charges was constructed. The main take away from this construction is that the multiparticle asymptotic states of the S-matrix are \textit{not} tensor products of single particle states. In particular, under Lorentz transformations, they pick up an extra little group phase for every monopole-charge (or dyon-dyon) pair. For example, consider a 2-particle state where each particle has electric and magnetic charges $(e_i,g_i)$ and spin $s_i$. This state transforms as
\begin{eqnarray}\label{eq:pairwisetransform}
&&U(\Lambda)\ket{p_{i},p_{j};s_{i},s_{j};q_{ij}}=\nonumber\\[5pt]
&&e^{iq_{ij}\phi_{ij}}\,\mathcal{D}_{s'_{i},s_{i}}\mathcal{D}_{s'_{j},s_{j}}\ket{\Lambda p_{i},\Lambda p_{j};s'_{i},s'_{j};q_{ij}}\,,
\end{eqnarray}
Here $U(\Lambda)$ is the unitary representation of the Lorentz transformation $\Lambda$, while the $\mathcal{D}_{ab}$ represent single particle little group factors. The extra ``pairwise little group" phase $e^{iq_{ij}\phi_{ij}}$ is unique to multiparticle states involving monopoles and charges (or any other mutually non-local particles). The pairwise helicity $q_{ij}$ is half-integer since it labels charges under the pairwise little group, which is a compact $U(1)$ \cite{Csaki:2020inw}. It has a natural interpretation as the quantity
\begin{eqnarray}\label{eq:q}
q_{ij}\,=\,e_{Mi} g_{Mj} - e_{Mj} g_{Mi}\,,
\end{eqnarray}
which is quantized in half integer units by the Dirac-Zwanziger-Schwinger quantization condition 
\cite{Dirac}.

A more detailed definition of electric magnetic multiparticle states was given in \cite{Csaki:2020yei}.
The transformation rule \eqref{eq:pairwisetransform} implies additional constraints on scattering amplitudes involving monopoles---the functional form of the scattering amplitude has to be such that 
\begin{eqnarray}\label{eq:pairwise}
&&\mathcal{A}(\Lambda p_1,\ldots,\Lambda p_n;\Lambda k_1,\ldots,\Lambda k_m)=\nonumber\\[5pt]
&&e^{-i\sum q_{ij}\phi_{ij}}\,\tilde{\mathcal{A}}(p_1,\ldots,p_n; k_1,\ldots,k_m)\,,
\end{eqnarray}
where $\tilde{\mathcal{A}}$ is the amplitude $\mathcal{A}$ times all of the single particle little group transformations $\mathcal{D}_i$. To construct amplitudes with the required transformation rule ref. \cite{Csaki:2020inw} defined new spinor-helicity variables called ``pairwise spinors," denoted by $\ket{p^{\flat\pm}_{ij}}$, defined for each pair of particles in the in or out state. For completeness, we repeat the definition of these spinors in the Appendix. The spinors have pairwise helicity $\pm$ under the pairwise little group associated with the particles $i$ and $j$. In other words, they transform as
\begin{eqnarray}\label{eq:transf}
{\tilde \Lambda} \ket{p^{\flat\pm}_{ij}}\,&=&\, e^{\pm \frac{i}{2}\phi(p_i,p_j,\Lambda)}\,\ket{\Lambda p^{\flat\pm}_{ij}}\nonumber\\[5pt]
\sbra{p^{\flat\pm}_{ij}} \,{\tilde \Lambda} \,&=&\,e^{\mp \frac{i}{2}\phi(p_i,p_j,\Lambda)}\,\sbra{\Lambda p^{\flat\pm}_{ij}}\, ,
\end{eqnarray}
where $\Lambda$ and $\tilde \Lambda$ are Lorentz transformations acting in vector and spinor spaces respectively. Finally, the pairwise spinors have the important property that they align with some of the standard spinor helicity variables in the massless limit. In particular:
\begin{eqnarray}\label{eq:vanish}
&&\bk{i\,p^{\flat+}_{ij}}= \bks{i\,p^{\flat+}_{ij}}=0\nonumber\\
&&\bk{j\,p^{\flat-}_{ij}}= \bks{j\,p^{\flat-}_{ij}}=0~.
\end{eqnarray}
The vanishing of these contractions plays a central role in explaining the peculiarities of the Rubakov-Callan effect.

To see the relation between pairwise helicity and the Rubakov-Callan effect, let us consider an incoming state involving the massless fermions $u^1,\,u^2$, both with electric charge $e_M=-1/2$ and a scalar monopole $M$ with magnetic charge $g_M=-1$. Let us now focus on the s-wave partial amplitude involving in- and out- states with total angular momentum $J=0$. In this case the amplitude splits into an incoming and and outgoing part, each one depending only on the incoming/outgoing momenta and with all spinor indices contracted (since $J=0$). As $q_{u^1,M}=q_{u^2,M}=-1/2$, the incoming part of the amplitude is
\begin{eqnarray}\label{eq:incrub}
\bks{u^1\, p^{\flat -}_{u^1,M}}\bks{u^2\, p^{\flat -}_{u^2,M}}\,,
\end{eqnarray}
where $\sket{p^{\flat -}_{u^i,M}}$ are pairwise spinors, while $\sbra{u^i}$ are the standard massless spinor helicity variables.
To see that this in-state transforms correctly, note that the $\sket{p^{\flat -}_{u^i,M}}$ each carry pairwise helicity $-1/2$ under the ${u^i,M}$ pairwise little group, while the $\sbra{u^i}$ transform like a helicity $1/2$ under the single particle little group for $u^i$, which is suitable since incoming left-handed fermions carry helicity $1/2$ in our \textit{all-outgoing} convention. In contrast, outgoing left-handed fermions carry helicity $-1/2$ in this convention. Note that pairwise helicity is not flipped between incoming and outgoing particles \cite{Csaki:2020inw}. 

We can now easily see why there can't be forward scattering in this process. Let us try to represent the would-be out-state relevant for forward scattering, i.e. involving the same $u^1,\,u^2$. The out part of the amplitude has to be 
\begin{eqnarray}\label{eq:outrubno}
\bk{u^1 \, p^{\flat +}_{u^1,M}}\bk{u^2\, p^{\flat +}_{u^2,M}}\,.
\end{eqnarray}
Note that the sign on the pairwise spinors is flipped so as to preserve their pairwise helicity under $\sket{}\rightarrow\ket{}$. However, this expression vanishes by \eqref{eq:vanish}. There cannot be forward scattering of fermions on a monopole in the lowest partial wave. 

Having established that there is no forward scattering for the Rubakov-Callan in-state, we now turn to write down the only possible final state which respects all SM quantum numbers, as well as the overall $SU(4)$ flavor symmetry. This out state involves the fermions $(\bar{e})^\dagger,(\bar{d}^3)^\dagger$. The corresponding outgoing part of the amplitude is
\begin{eqnarray}\label{eq:outrub}
\bks{\bar{e}^\dagger\, p^{\flat -}_{\bar{e}^\dagger,M}}\bks{\bar{d}^{3\dagger} \,p^{\flat -}_{\bar{d}^{3\dagger},M}}\,.
\end{eqnarray}
It transforms correctly under the pairwise little group, since $q_{\bar{e},M}=q_{\bar{d}^3,M}=1/2$. Since this is the only possible out state, we have a simple derivation of the Rubakov-Callan amplitude
\begin{eqnarray}\label{eq:allrub}
&&\mathcal{A}_{\text{Rubakov-Callan}}\propto\nonumber\\[5pt]
&&\bks{u^1\, p^{\flat -}_{u^1,M}}\bks{u^2 \,p^{\flat -}_{u^2,M}}\bks{\bar{e}^\dagger\, p^{\flat -}_{\bar{e}^\dagger,M}}\bks{\bar{d}^{3\dagger}\, p^{\flat -}_{\bar{d}^{3\dagger},M}}\,.
\end{eqnarray}
The overall cross section for the process satisfies the s-wave unitarity bound, and so should be proportional to $4\pi p^{-2}_c$ where $p_c$ is the COM momentum. When taking QCD confinement of the incoming quarks into account, the incoming quarks are confined to within a distance $\Lambda^{-1}$ of each other, and the cross section becomes $\mathcal{O}(\Lambda^{-2})$.

\section{Solving a 40 Year old Mystery}
When a positron, $\bar{e}$, scatters off of a GUT monopole, forward scattering is again forbidden by angular momentum conservation, while the flavor symmetry constrains the out state to have 3 (mod 4) fermions. The only possible out state with 3 fermions which conserves all quantum numbers is:
\beq\label{e:OurOut}
\bar{u}^{1\dagger}\,+\,\bar{u}^{2\dagger}\,+\,{\bar d}^{3\dagger}\,.
\eeq
However, Callan argued that this final state is impossible, since in the presence of the monopole, the $\bar{d}^{3\dagger}$ cannot exist in a one-particle outgoing partial wave with $J=0$. Working in a truncated 2D theory including only fermions in one-particle $J=0$ waves which are then bosonized to solitons, Callan found that the final state consists of four semitons, or ``half-particles". For the initial state of an ${\bar e}$ he found the semitonic final state $1/2( e^\dagger + {\bar u}^{1\dagger}+  {\bar u}^{2\dagger} + d^3)$. Since ``half-particles" do not exist in the 4D theory Callan suggested the interpretation that half the time one would produce a positron and half the time one would produce a proton. These proposed individual processes do not conserve SM gauge charges, but would do so on average.

Analog 2D theories with an $SO(8)$ global symmetry have been analyzed by Maldacena and Ludwig \cite{Maldacena:1995pq} and Boyle Smith and Tong \cite{Smith:2020nuf}. 
These authors confirmed that in the absence of additional gauge symmetries the semiton description is correct, and can be understood via $SO(8)$ triality. However this does not answer the question of what happens for the GUT monopole process where the fermions have chiral non-Abelian charges that break the $SO(8)$ symmetry.
 
Sen \cite{Sen} claimed that conservation laws ensure that there are no monopole processes allowed with one fermion in the initial state and three fermions in the final state. If this were true then there would either have to be processes with more fermions (3 mod 4) or a mechanism that prevented single fermions from encountering a monopole.  However the conservation laws that Sen used are only valid in the 2D truncated theory which leaves out the possibility of entangling a fermion with the field angular momentum produced by a different particle, so it is not surprising that his analysis cannot produce the correct final state.

Kitano and Matsudo \cite{Kitano:2021pwt} suggested that the semitons should be identified in the 4D theory with a ``pancake" soliton: a domain wall bounded by a string. These pancakes are supposed to be heretofore unknown asymptotic states of the gauge theory. For this to be a consistent interpretation in the massless fermion limit, the pancake would also have to have arbitrarily small energies since the incoming positron energy can be arbitrarily small.

Using the pairwise helicity formalism, we are able for the first time to identify the correct final state for positron-monopole scattering. This final state does, in fact, consist of the fermions in \eqref{e:OurOut}, which conserve all of the SM quantum numbers and respect the approximate $SU(4)$ flavor symmetry. The novelty here is that the final state fermions are in fact \textit{entangled} with the field angular momentum arising from one of the other particles, which allows them to be in an overall $J=0$ state, even though they are not in one-particle $J=0$ states. The amplitude for this process is
\begin{eqnarray}\label{eq:poistron}
&&\mathcal{A}_{\bar{e}}\propto\nonumber\\[5pt]
&&\bks{\bar{e}\, p^{\flat -}_{\bar{e}^\dagger,M}}\bks{\bar{u}^{1\dagger}\, p^{\flat -}_{\bar{u}^{1\dagger},M}}\bks{\bar{u}^{2\dagger}\, p^{\flat +}_{\bar{d}^{3\dagger},M}}\bks{\bar{d}^{3\dagger}\, p^{\flat -}_{\bar{u}^{2\dagger},M}}\nonumber\\[3pt]
&&\,-\,(1\leftrightarrow 2)\,.\nonumber\\
\end{eqnarray}

Note that we cannot arrange a similar cross-entanglement when there are only two fermions in the final state. Consider the static monopole limit, then in the center of mass frame the two fermions emerge back-to-back, and the flat momenta are also back-to-back along this axis, thus exchanging the flat spinors so that they are contracted with the opposite particle gives exactly zero. For finite monopole masses there could be a contribution that is suppressed by the monopole mass, but this cannot saturate the unitarity bound. Also note that truncating to 2D also forces all the momenta to be along a single direction, so at least one exchange of flat momenta will give a vanishing amplitude. Thus we can easily see how the 2D truncation fails to capture the 4D physics. In the static limit the three quarks are coplanar in the center of momentum frame, so there is no obstruction to entanglement in a 3D theory.

\section{Applications}
 Since cross sections that saturate partial wave unitarity grow with the inverse of the initial momentum one might naively expect that the positron scattering process we have discussed would lead to an arbitrarily large cross section for B-violation in GUT theories. We can see however that the growth is cut off at the QCD scale, as happens for the Rubakov-Callan processes.  Once the initial energy is below the sum of the monopole and proton masses, the final state of three quarks cannot hadronize into a proton.  In the monopole rest frame the three quark state will carry the initial momentum of the positron, and once the separations of the quarks reaches the QCD scale the quarks will be forced to travel in the same direction, so two of the quarks will have their momentum flipped by QCD interactions.  Since QCD also breaks chirality, their chirality can also be flipped and they can become in-states for a second interaction with the monopole.  Two quarks scattering on the monopole produce an antiquark and a lepton.  The antiquark can annihilate with the remaining quark to produce two photons or a lepton-antilepton pair. Thus below the proton threshold there is no B-violation, as we expect from energy conservation, and the B-violating cross section is cut off at the QCD scale.

\section{Acknowledgments}

\begin{acknowledgments}
We thank Sungwoo Hong,  Ryuichiro Kitano, Juan Maldacena, Ryutaro Matsudo, Joe Polchinski, and David Tong for helpful conversations.
CC is supported in part by the NSF grant PHY-2014071 as well as the US-Israeli BSF grant 2016153.  OT was supported in part by the DOE under grant DE-AC02-05CH11231.
Y.S. is supported in part by the NSF grant  PHY-1915005.
J.T. is supported by the DOE under grant  DE-SC-0009999. We thank the Aspen Center for Physics, which is supported by National Science Foundation grant PHY-1607611, where parts of this work were completed.
\end{acknowledgments}

\section{Appendix}
In this appendix we repeat for completeness the definitions of pairwise spinors from \cite{Csaki:2020inw}. For every pair $(i,j)$ of particles, we can easliy define the pairwise spinors in the COM frame of the pair, where the momenta of the two particles are
\begin{eqnarray}
k^i=(E_i,p_c\,\hat{z}),~\,k^j=(E_j,-p_c\,\hat{z}),
\end{eqnarray}
and $E_{i,j}=\sqrt{m_{i,j}+p^2_c}$. As a first step we define the two null momenta:
\begin{eqnarray}
k^{\flat\pm}_{ij}=p_c\,(1,\pm\hat{z})\,.
\end{eqnarray}
It's easy to see that they are linear combinations of $k_i$ and $k_j$. Our pairwise spinors in the COM frame are simply the ''square roots" of these pairwise momenta:
\begin{eqnarray}
&&\ket{k^{\flat+}_{ij}}_{\alpha}=\sqrt{2\,p_c}\,\colvec{1\\ 0}\,,\,\ket{k^{\flat-}_{ij}}_{\alpha}=\sqrt{2\,p_c}\,\colvec{0\\1}\nonumber\\
&&\sbra{k^{\flat+}_{ij}}_{\dot{\alpha}}=\sqrt{2\,p_c}\,\left(1~ ~0\right)\,,\,\sbra{k^{\flat-}_{ij}}_{\dot{\alpha}}=\sqrt{2\,p_c}\,\left(0~ ~1\right)\, .\nonumber\\
 \end{eqnarray}
These are defined so that  
 \begin{eqnarray}
k^{\flat\pm}_{ij} \cdot \sigma_{\alpha{\dot{\alpha}}}=\ket{k^{\flat\pm}_{ij}}_{\alpha}\sbra{k^{\flat\pm}_{ij}}_{\dot{\alpha}}\, .
 \end{eqnarray}
To get the spinors in any other frame with momenta $p_{i,j}=L_p k_{i,j}$ (also $p^{\flat \pm}_{ij}=L_{p} k^{\flat \pm}_{ij}$), we simply act on them with the Lorentz transformation $L_{p}$ in the spinor representation, denoted by $\mathcal{L}_p,\,\tilde{\mathcal{L}}_p$. We define
 \begin{eqnarray}
\ket{p^{\flat\pm}_{ij}}_{\alpha}&=&\left(\mathcal{L}_p \right)^{~\beta}_{\alpha}\,\ket{k^{\flat\pm}_{ij}}_\beta\nonumber\\
\sbra{p^{\flat\pm}_{ij}}_{\dot{\alpha}}&=&\sbra{k^{\flat\pm}_{ij}}_{\dot{\beta}}\,\left( \tilde{\mathcal{L}}_p \right)^{\dot{\beta}}_{~\dot{\alpha}}.
 \end{eqnarray}
This guarantees the relation
 \begin{eqnarray}
p^{\flat\pm}_{ij} \cdot \sigma_{\alpha{\dot{\alpha}}}=\ket{p^{\flat\pm}_{ij}}_{\alpha}\sbra{p^{\flat\pm}_{ij}}_{\dot{\alpha}}\, .
 \end{eqnarray}
In \cite{Csaki:2020inw} it was shown that the pairwise spinors $\ket{p^{\flat\pm}_{ij}},\,\sket{p^{\flat\pm}_{ij}}$ defined above transform with the correct pairwise little group phase
\begin{eqnarray}\label{eq:transf}
\Lambda^{~\beta}_{\alpha}\ket{p^{\flat\pm}_{ij}}_{\beta}&=&e^{\pm \frac{i}{2}\phi(p_i,p_j,\Lambda)}\,\ket{\Lambda p^{\flat\pm}_{ij}}_{\alpha}\nonumber\\[5pt]
\sbra{p^{\flat\pm}_{ij}}_{\dot{\beta}} \,\tilde{\Lambda}^{\dot{\beta}}_{~\dot{\alpha}}&=&e^{\mp \frac{i}{2}\phi(p_i,p_j,\Lambda)}\,\sbra{\Lambda p^{\flat\pm}_{ij}}_{\dot{\alpha}}.\nonumber\\
 \end{eqnarray}

\end{document}